\documentclass[journal,letterpaper]{IEEEtran_nssmic}
\usepackage{cite}       
\usepackage{graphicx}   
\usepackage{amsmath}    
\usepackage{amssymb}
\usepackage{array}

\begin{document}

\renewcommand{\labelenumi}{\arabic{enumi}}
\renewcommand{\labelitemi}{-}

\title{Test Beam Characterizations of 3D Silicon Pixel Detectors}
\author{M.~Mathes, M.~Cristinziani, C.~Da~Via', M.~Garcia-Sciveres, K.~Einsweiler, J.~Hasi,\\
C.~Kenney, Sherwood~Parker, L.~Reuen, M.~Ruspa,
J.~Velthuis, S.~Watts, N.~Wermes\\
\thanks{Manuscript received June 01, 2008}
\thanks{The corresponding authors Markus Mathes and Norbert Wermes are with Bonn University, Physikalisches Institut,
Nussallee 12, D-53115 Bonn, Germany, email: mathes@physik.uni-bonn.de, wermes@uni-bonn.de}
\thanks{M.~Cristinziani and L.~Reuen are with Bonn University, Physikalisches Institut, Nussallee 12, D-53115 Bonn, Germany.}
\thanks{J.~Velthuis was with Bonn University and is now with University of Bristol, Tyndall Avenue, Bristol BS8 1TL, UK.}
\thanks{C.~Da~Via, J.~Hasi, S.~Watts are with Manchester University, Oxford Road, Manchester, M13 9PL, UK.}
\thanks{S.~Parker is with the Univ. of Hawaii, Honolulu, HI, 96822, USA}
\thanks{M.~Ruspa is with the Univ. Piemonte Orientale, Novara and INFN Torino, Italy.}
\thanks{C.~Kenney is with the Molecular Biology Consortium, California, USA}
\thanks{K.~Einsweiler, M.~Garcia~Sciveres are with Lawrence Berkeley National Laboratory, 1 Cyclotron Road,
Berkeley, CA 94720, USA}
\thanks{Work supported by the German Ministerium f{\"u}r Bildung,
              Wissenschaft, Forschung und Technologie (BMBF) under contract
              no.~06 HA6PD1 and by the
              U.S. Department of Energy Office of Science under Contract No. DE-AC02-05CH11231.}
}

\maketitle

\begin{abstract}
3D silicon detectors are characterized by cylindrical electrodes
perpendicular to the surface and penetrating into the bulk material in contrast to standard Si detectors with
planar electrodes on its top and bottom. This geometry renders
them particularly interesting to be used in environments where
standard silicon detectors have limitations, such as for example the
radiation environment expected in an LHC upgrade. For the first time,
several 3D sensors were assembled as hybrid pixel detectors using
the ATLAS-pixel front-end chip and readout electronics. Devices with
different electrode configurations have been characterized in a 100 GeV pion
beam at the CERN SPS. Here we report
results on unirradiated devices with three 3D electrodes per 50$\times$400 $\mu$m$^2$ pixel
area.
Full charge collection is obtained already with comparatively low bias voltages
around 10 V.
Spatial resolution with binary readout is obtained as expected from
the cell dimensions. Efficiencies of $95.9 \% \pm
0.1 \%$ for tracks parallel to the electrodes and of $99.9 \% \pm 0.1 \%$ at
15$^\circ$ are measured. The homogeneity of the efficiency over
the pixel area and charge sharing are characterized.
\end{abstract}


\section{Introduction}
The silicon tracking devices at the LHC detectors ATLAS and CMS have
to cope with very intense radiation levels reaching fluence levels of
up to 10$^{15}$ particles per cm$^{2}$ in the innermost pixel layer
during a ten year LHC projected life time~\cite{LHC-fluences}. This
large particle fluence causes damage to silicon pixel sensors, most
dominantly by: (a) the trapping of moving charges which reduces the
signal, (b) an increase in the negative space charge concentration
N$_{\rm eff}$ of the depleted silicon bulk which needs a larger
external bias voltage to reach full
depletion, and (c) an increase of leakage currents due to new energy
states inside the band gap. The semiconductor tracking detectors of
the large LHC experiments ATLAS and CMS have been designed and built
to stand these challenges for the LHC life
time~\cite{ATLAS-TDR,CMS-TDR,ATLAS_Sensor,CMS-pixel}. As plans for an
upgrade of the LHC which may eventually lead to an increase in
luminosity by a factor of 10 (sLHC, Super-LHC) are
discussed~\cite{sLHC}, new types of sensors for pixel detectors,
either by material or by design, are being investigated to be able to
cope with the concurrent increase also in radiation fluence of up to
10$^{16}$ particles per cm$^{2}$. For such radiation levels the
present detectors are not suited. So-called 3D-silicon sensors have
been proposed and developed (\cite{3D_1997,kok2006} and references therein). Due to a different
electrode configuration 3D sensors are more radiation hard and have
faster charge collection than standard planar silicon sensors.
They are thus a prime candidate for pixel sensors at the sLHC.

\section{The tested structures}\label{structures}
\begin{figure}[thb]
\begin{center}
\includegraphics[width=0.4\textwidth]{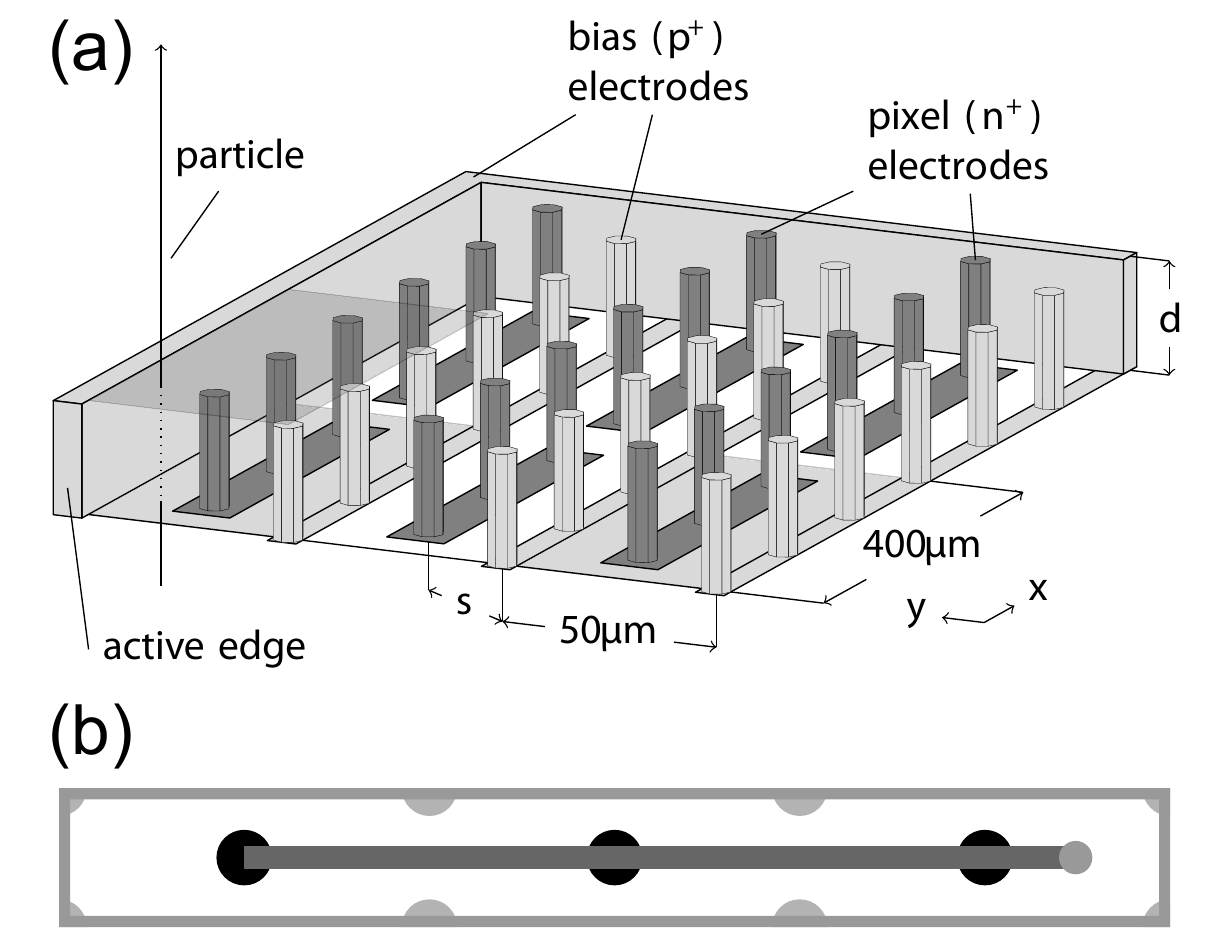}
\end{center}
\caption[]{\label{3D-silicon principle} (a) Schematic view of 3D silicon structures.
3D view showing the p and n electrodes and the active edge; the distance s is 70.6 $\mu$m. (b) view from top of one pixel cell (bump bond pad shown at far right).}
\end{figure}

The basic concept of 3D silicon sensors is shown in
fig.~\ref{3D-silicon principle}. A 3-dimensional structure is obtained
by processing the n$^+$ and p$^+$ electrodes into the substrate bulk by
combining VLSI and MEMS (Micro Electro Mechanical Systems)
technologies~\cite{3D-manufacturing}. Charge carriers drift
inside the bulk parallel to the surface over a typical drift
distance of 50 -- 100 $\mu$m. Typical depletion voltages are of
the order of 10 V.

In this paper we have particularly studied the response of 3D pixel
devices to high energy particles in terms of
response homogeneity and spatial resolution. A wafer with three different types of 3D
structures geometrically adapted to fit the pixel pattern of the ATLAS
pixel frontend chip FE-I3~\cite{FEI3} has been produced at the
Stanford Nanofabrication Facility~\cite{stanford-microlab}. The thickness of the p-type sensors is
208 $\pm$5 $\mu$m. The nominal etched diameter of the 3D-columns is 17$\mu$m.
PbSn underbump deposition
and flip-chipping to bumped FE-I3 chips was done at
IZM-Berlin~\cite{IZM1,IZM2}. The pixel area of the FE-I3 chip is
50$\mu$m$\times$400$\mu$m. Different structures with different distances
between the 3D electrodes were produced to fit the FE-I3 pixel
dimensions with two (2E), three (3E), or four (4E) equally-spaced electrodes in the
area of one electronics pixel. In this paper we report on the study of the
configuration with three electrodes (3E, Fig.\ref{3D-silicon principle}). An evaluation of the 2E, 3E, and 4E
structures against each other will be the subject of a future publication. The tested devices were still unirradiated.
The ATLAS pixel electronics provides zero
suppression in the readout with a threshold level for this study of about 3000e$^-$.
The electronics noise is about 380e$^-$. Analog information is obtained via the in-pixel measurement of
time-over-threshold (ToT) with an approximate resolution of 7~bits~\cite{FEI3}.
The calibration is approximately linear for signal charges above 5000\,e$^-$ with one ToT unit corresponding to about
400\,e$^-$.

\section{Test beam setup and environment}\label{testbeam}
\begin{figure}[thb]
\begin{center}
\includegraphics[width=0.5\textwidth]{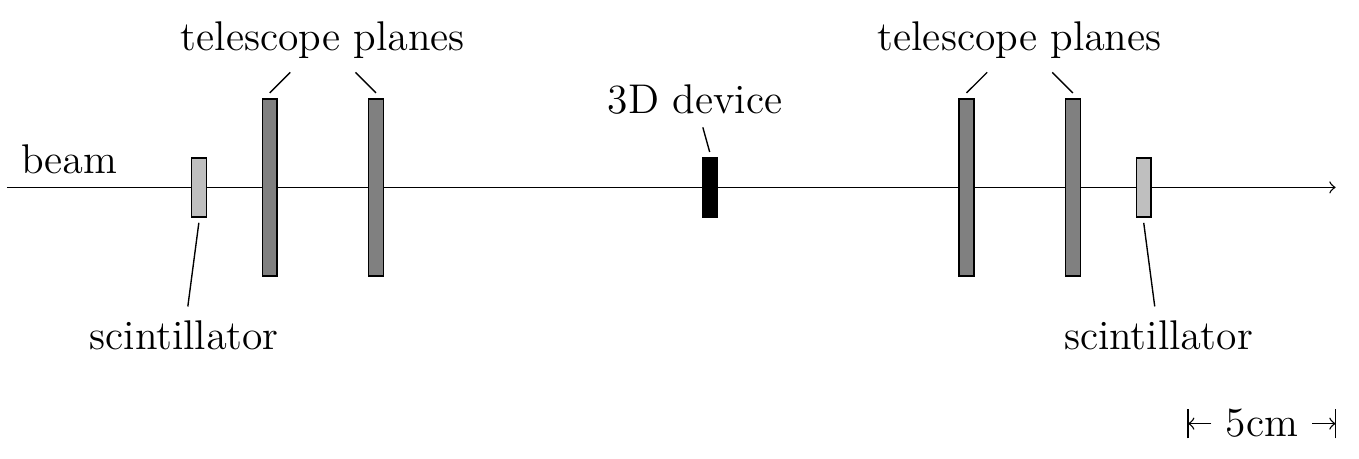}
\caption[]{Reference telescope and position of the tested 3D structures in the CERN 100 GeV pion test beam.}
\label{testbeam-setup}
\end{center}
\end{figure}

The 3D silicon devices have been tested in a 100 GeV beam of pions at
the CERN SPS. The setup is shown schematically in
fig.~\ref{testbeam-setup}. The device under test (DUT, 3D silicon
pixels) was placed in between two pairs of silicon microstrip
detectors~\cite{beam_telescope}. The telescope was developed for ATLAS
and consists of double sided silicon microstrip detectors with
50$\mu$m pitched strips on both sides rotated by 90$^\circ$ with
respect to each other. The S/N ratio of about 37 for the p-side,
measuring the DUT x-direction, is better than for the n-side (S/N
$\approx$ 22), measuring the DUT y-direction. Hits are read out zero
suppressed and events waiting for read out can be buffered for some
time~\cite{beam_telescope}. The latter option was, however, not
used. The setup is triggered by the coincidence signal of two
scintillators in front of and behind the setup. The precision obtained in
the plane of the DUT is better than 5.3~$\mu$m in both spatial
directions.
The beam divergence is measured to be less than 0.2~mrad. The beam incidence angle with respect to the DUT plane
is at most 1$^\circ$.
The average data taking rate was limited by the readout system to
around 50-60 Hz. The results presented in this paper concentrate on 3D
devices with three electrodes. Bias voltages between 6V and 25V in
steps of 4V were applied. Two angles of inclination with respect to
the beam, 0$^\circ$ and 15$^\circ$, were studied.
Only tracks from events with a single hit in each of the telescope
planes were selected for characterization measurements.
\emph{Hits} in the 3D-device were accepted if the seed position
of the cluster is found in the predicted or in a neighboring pixel.
Otherwise they are counted as \emph{misses}.

\section{Results}\label{results}
\begin{figure}[thb]
\begin{center}
\includegraphics[width=0.5\textwidth]{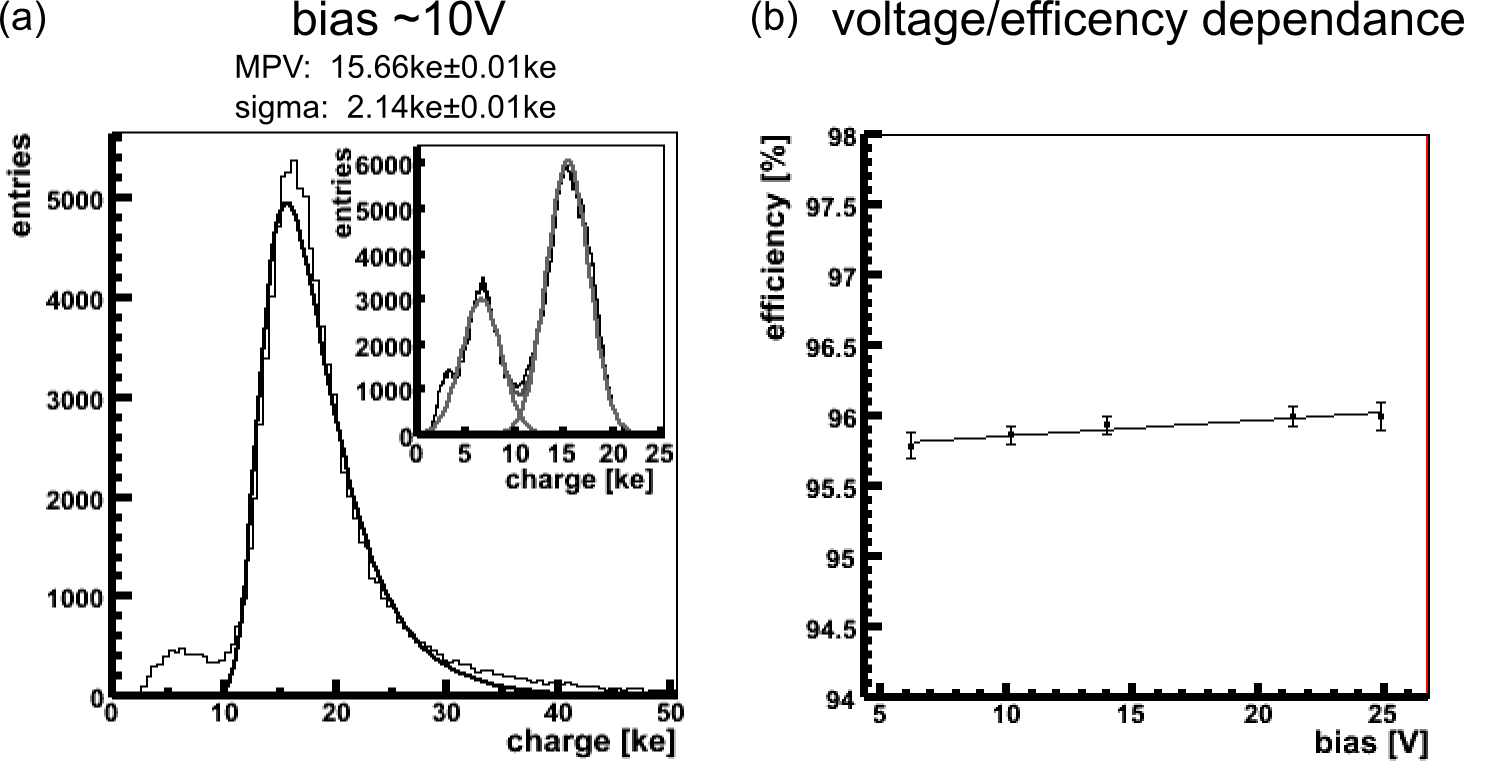}
\caption[]{(a) Charge distribution of perpendicularly incident particles measured using a bias voltage of 25\,V. The solid line is a Landau fit.
The quoted errors on mean and sigma are obtained by the fit. The insert shows the $^{241}$Am lines of 59.6 keV and 26.3 keV, respectively. The calibration uncertainty is between 5$\%$ and 9$\%$.
(b) Hit efficiency as a function of applied bias voltage from 6.5\,V to 31.5\,V.}
\label{charge-collection}
\end{center}
\end{figure}

\begin{figure}[thb]
\begin{center}
\includegraphics[width=0.5\textwidth]{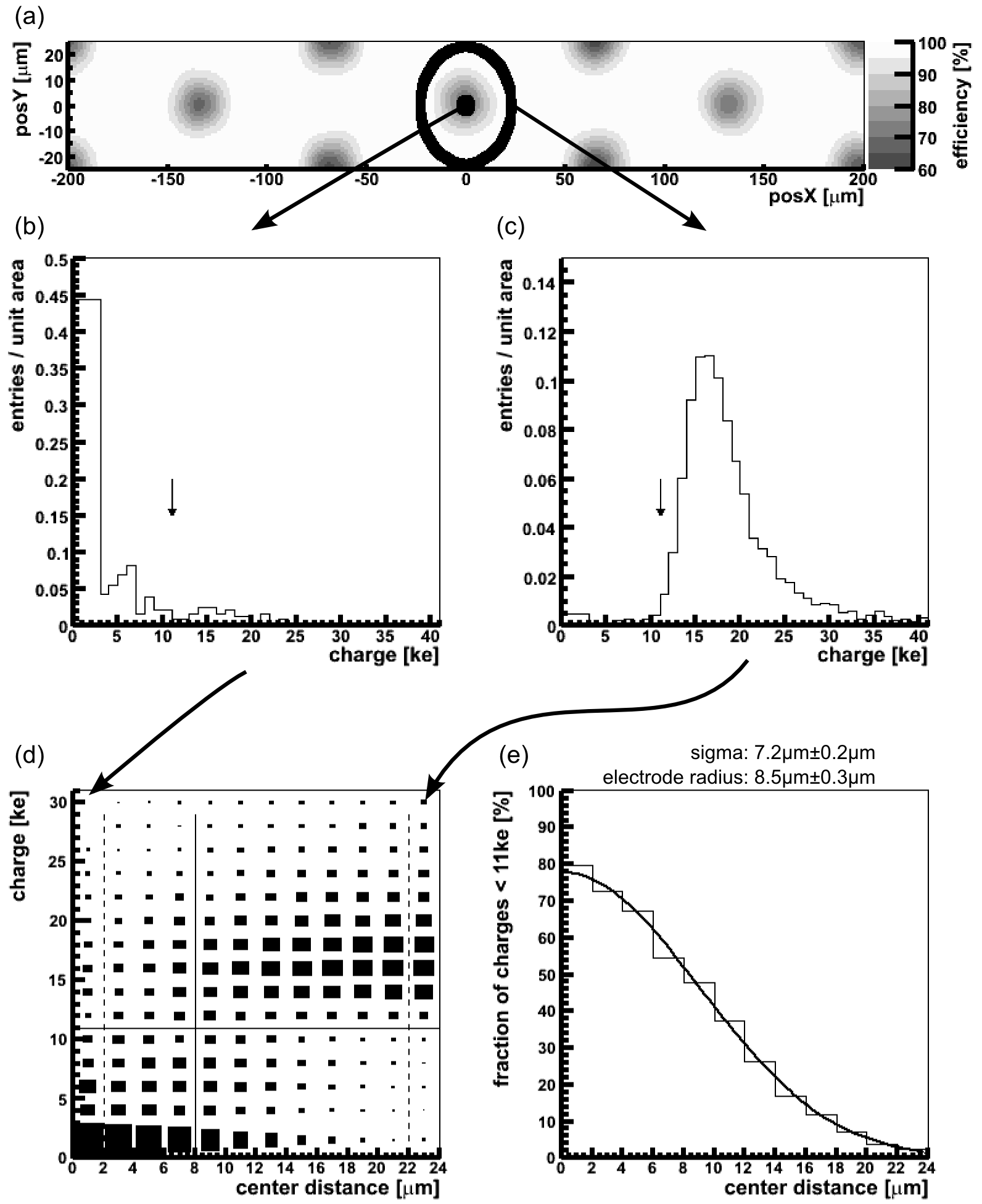}
\caption[]{(a) Charge collection map over the area of one pixel
  (50$\times$400$\mu$m). The areas indicated by the black dot and the ring lead to charge distributions shown in (b) and (c) below,
  representing areas in and far away from the 3D electrodes, respectively. The arrows in (b) and (c) indicate the chosen separation selection for the two-dimensional histogram in (d) showing charge versus
  distance from the column center. The lowest and the highest x-bins in (d) correspond to the charge distributions in (b) and (c) as indicated.
  (e) Relative fraction of entries per area with a charge smaller that 11\,ke$^-$. The superimposed curve is a fit (see text).}
\label{charge-losses}
\end{center}
\end{figure}

\paragraph*{Operation characteristics}
3D devices can be fully depleted with low bias voltages compared to
planar silicon sensors. The theoretically expected voltage for
cluster signal saturation is about 8\,V. In
fig.~\ref{charge-collection}(a) the measured charge distribution with a Landau type shape is
shown as obtained by operating the DUT with a bias voltage of 10\,V.
The insert shows the 59.6 keV and 26.3 keV $\gamma$-lines from $^{241}$Am, indicating that the ToT calibration uncertainty lies
between 5-9$\%$. From the fit, the values for peak, mean and sigma
for the fully depleted device are 15700~e$^-$, 19200~e$^-$ and
2100~e$^-$, respectively, the latter being mostly due to Landau fluctuations.
The population of entries at small
charges on the left hand side of fig.~\ref{charge-collection}(a) indicate
charge losses when tracks pass directly through the 3D-electrodes. This will be studied in more detail below.
For data taking the 3D devices were operated with several
different bias voltages. We show in fig.~\ref{charge-collection}(b) the hit
efficiency, obtained by integrating all hits and misses over the area of a complete pixel cell,
as a function of the applied bias voltage. The efficiency is almost unchanged even down to voltages of $\sim$6\,V.
Data using even lower bias voltage were not taken during the test beam.
The slight increase with bias voltage seen in fig.~\ref{charge-collection}(b) we interpret as coming from
effects due to the higher electric fields inside the sensor rather than being due to a change in the depletion volume.
%
\paragraph*{Hit efficiency}
In order to investigate the spatial dependence of the hit efficiency,
the sensors with three electrodes per pixel have been scanned with the
beam to obtain a hit efficiency map. Scans with straight tracks
(perpendicular incidence) and with tracks under an inclination angle
of about 15$^\circ$ were carried out. For straight track, the obtained efficiency map is shown in fig.~\ref{charge-losses}(a).
Tracks from all illuminated pixels enter the efficiency map.
The spot-resolution with which the map is drawn is given by the 2-dimensional bin size,
smoothed by a gaussian, folded with the telescope extrapolation
error. All combined this corresponds to $\sim$6.5\,$\mu$m in both coordinates.
The position of the 3D electrodes
are clearly identified. The overall efficiency for perpendicularly
incident tracks, using a threshold setting of 3000\,e$^-$ and obtained
by integrating over the pixel area, corresponds to
\begin{eqnarray}\label{efficiency_straight}
\epsilon_{0^\circ} = 95.9 \% \pm 0.1 \% \qquad .
\end{eqnarray}
The respective cluster charge distribution from the two areas corresponding to low
(black central dot in fig.~\ref{charge-losses}(a) and high (ring between radii of 22$\mu$m and 24$\mu$m)
hit efficiency, are shown separately in figs.~\ref{charge-losses}(b) and (c), properly normalized by area.
Note that in fig~\ref{charge-losses}(b) the \emph{misses} are placed in the first bin and entries at and even somewhat below
the nominal threshold of 3000 e$^-$ are possible due to the non-linearity of the ToT calibration at small charges.
The charge distribution for tracks entering far away from the column center (fig.~\ref{charge-losses}(c))
shows the expected Landau shape with the lowest entries at about 10\,ke. Charge
entries below 11\,ke are thus classified as being due to tracks hitting the region
of the 3D electrode.
\begin{figure}[thb]
\begin{center}
\includegraphics[width=0.5\textwidth]{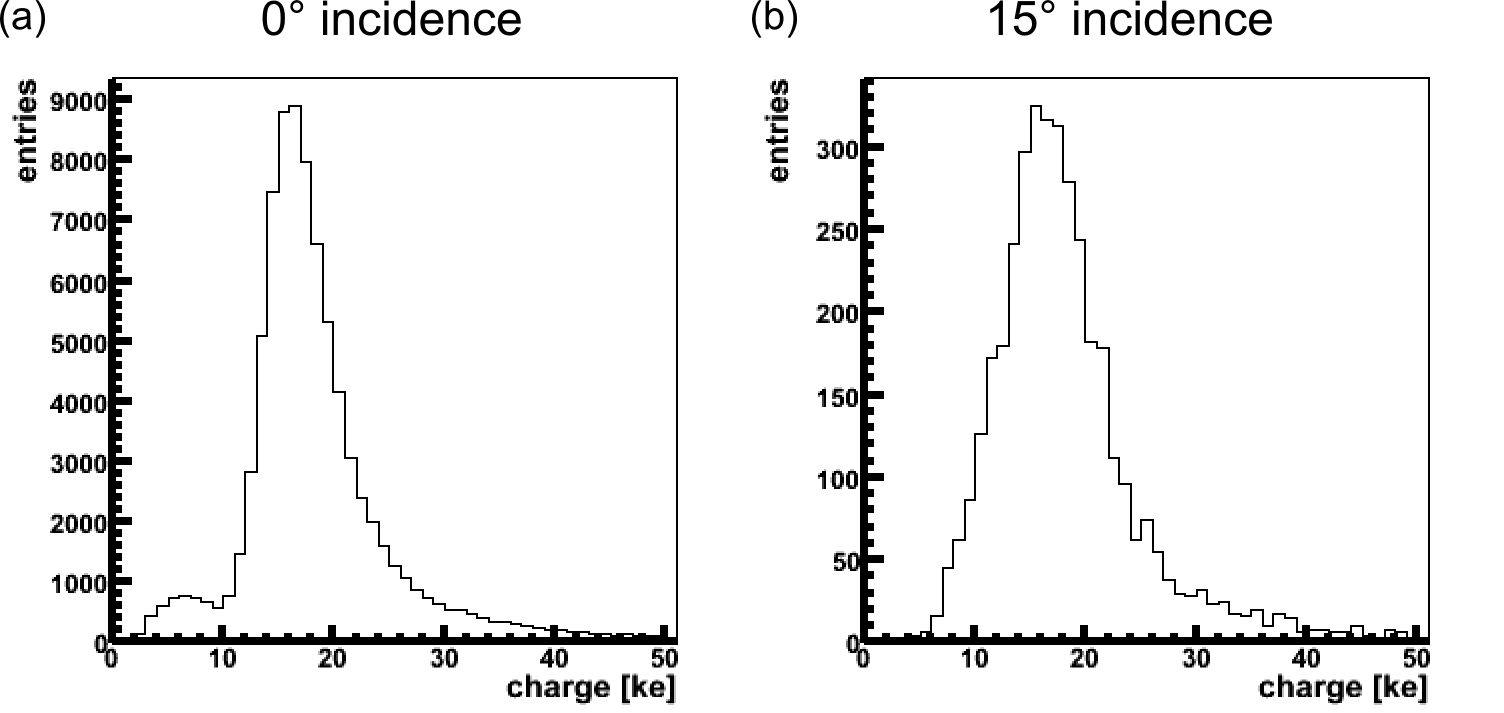}
\caption[]{Measured distributions of cluster charge for tracks under different
  inclination angles: (a) 0$^\circ$ angle of incidence, (b) 15$^\circ$ angle of
  incidence.}
\label{efficiency-15}
\end{center}
\end{figure}
In order to assess the region of reduced efficiency in the
3D-electrodes we plot in fig.~\ref{charge-losses}(d) the measured
charge versus the distance $\Delta$R from the nominal center of an
electrode column in a 2-dimensional histogram.
The solid lines separate regions above and below the low side of the Landau distribution
(arrows in fig.~\ref{charge-losses}(c) and (b)), as well as above and below the nominal column etched radius
(histogram bin starting at 8\,$\mu$m). In an ideal world this
plot would give a discrete charge output (a horizontal line) as a function of $\Delta R$.  This
ideal picture in reality is smeared by (i) the track extrapolation
resolution of about 5\,$\mu$m, (ii) the Landau distribution of deposited
energy in the sensor, (iii) $\delta$-electrons being mostly emitted at
$\sim$ 90$^\circ$ to the tracks and traveling into regions of more/less
efficient charge collection, and (iv) the electronics thresholds
of $\sim$ 3000\,e$^-$ existing for every pixel. Figure~\ref{charge-losses}(d) clearly
identifies two regions, one with large collected charge at distances larger than about 8\,$\mu$m from the electrode center
(upper right), and one with small charge at small distances, corresponding to the column region (bottom left).
Figure~\ref{charge-losses}(e), finally, displays the relative fraction of entries in rings around the electrode with a charge smaller that 11\,ke,
as a function of the distance from the electrode center. The superimposed curve represents a fit of a flat distribution with electrode radius~r
folded by a Gaussian. In agreement with a simple simulation one
finds a column radius of r = (8.5$\pm$0.3)\,$\mu$m, in agreement with the with of the nominal etching radius of 8.5\,$\mu$m.
The Gaussian $\sigma$ is found to be (7.2$\pm$0.2)~$\mu$m, i.e. wider than the contribution from the telescope resolution ($\sim$5$\mu$m) alone.
Part of this is due to the beam's $\lesssim$1$^\circ$ inclination to the DUT plane, resulting in a 4$\mu$m space displacement over the depth, but also
indicates that the efficiency inside the column is not a step function.
The additional spreading we hence interpret as being due to effects like charge diffusion or
other mechanisms of signal induction of tracks entering the undepleted column regions.

The mean charge of hits belonging to the
region outside of the electrode is about 20 -- 25\,ke. Hits assigned to
the electrode center, on the contrary, show a mean charge below 5ke.
Thus, with some uncertainty and caution introduced by the existing
pixel thresholds of $\sim$3\,ke, we estimate the charge
collection efficiency inside the 3D-electrode in the order
of at most 25\%. If, alternatively, one assumes 0$\%$ efficiency inside the electrode and 100$\%$ outside,
the \emph{effective column radius} is calculated from eq.~(\ref{efficiency_straight}) to be 6.5$\mu$m.

\begin{figure}[htb]
\begin{center}
\includegraphics[width=0.5\textwidth]{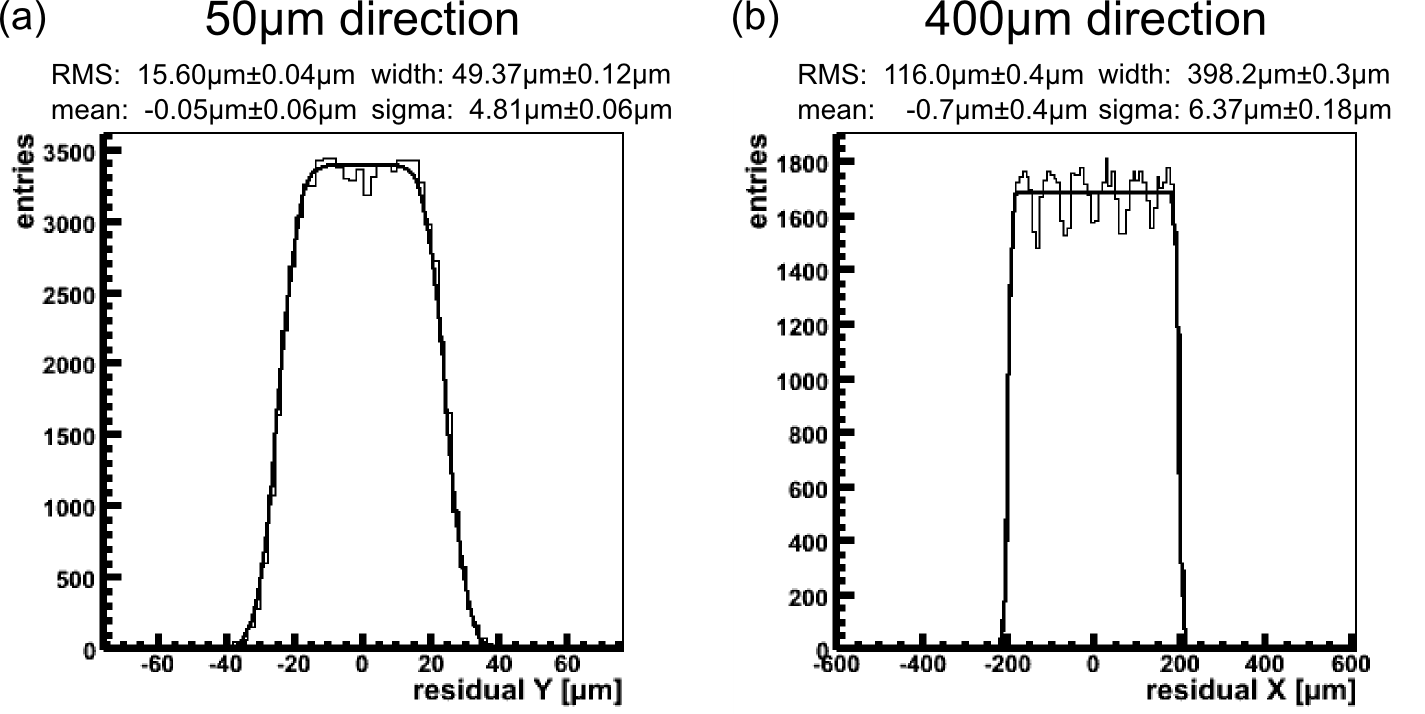}
\caption[]{Spatial resolution of the 3D pixel device measured with
  respect to the reference telescope. Only digital information is
  used, without interpolation using charge sharing. Plotted is the
  difference between the telescope space point and the measured space
  point. (a) resolution in the 50$\mu$m direction, (b) resolution in
  the 400$\mu$m direction. The distributions were fitted with a
  rectangular function convoluted by an Gaussian distribution. The
  stated width and sigma and their errors are the values obtained by
  the fit.}
\label{resolution}
\end{center}
\end{figure}

\begin{figure}[thb]
\begin{center}
\includegraphics[width=0.5\textwidth]{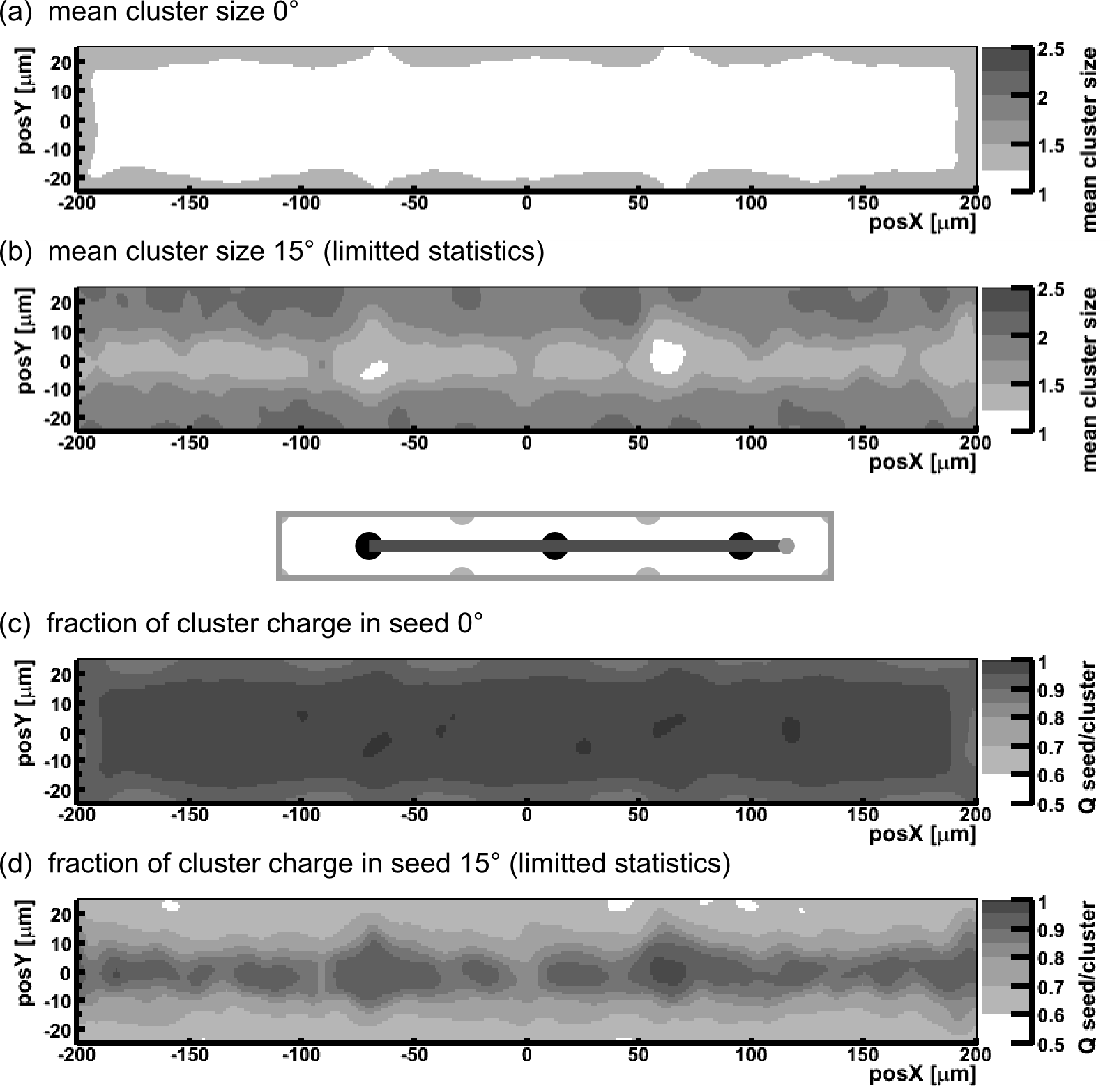}
\caption[]{Maps of (a),(b) the mean value of the hit-cluster size
  distribution, and (c),(d) the fraction of the charge of a
  hit-cluster seen in the pixel with the largest signal (seed
  pixel). The maps are for normally incident tracks (a)+(c) and
  for inclined tracks (b)+(d), 15$^\circ$ inclination, respectively.}
\label{charge-sharging}
\end{center}
\end{figure}

In a pixel tracking detector, the pixel modules are usually tilted in
the azimuthal direction to optimize charge sharing between neighboring
readout cells. For 3D sensors this produces the extra beneficial
effect that tracks originating from the interaction point would not
run exactly parallel through the center of the 3D electrodes. In
fig.~\ref{efficiency-15} Landau distributions for
perpendicular tracks (0$^\circ$) and for tracks under an inclination angle of 15$^\circ$ are compared.
As expected, for
inclined tracks in fig~\ref{efficiency-15}(b) the entries with small charge on the left of
fig.\ref{efficiency-15}(a) disappear. The width of the Landau
distribution now broadens by about 20$\%$ and shows the lowest detected
charge at about 5\,ke. While this is beneficial for track detection at collider detectors
where tracks always impinge the detector under an angle, the Landau broadening with entries as low as 5\,ke
does not constitute a comfortable distance to the threshold, in particular when a decrease of the signal charge
due to radiation must be expected. The hit efficiency for 15$^\circ$ inclined tracks
is measured to be
\begin{eqnarray*}
\epsilon_{15^\circ} = 99.9 \% \pm 0.1 \% \qquad .
\end{eqnarray*}
%
\paragraph*{Spatial resolution}
The spatial resolution is obtained by plotting the difference between
the track position predicted by the telescope on the plane of the DUT
(the 3D pixel sensor) and the reconstructed hit of the DUT
device. Figure~\ref{resolution} shows this distribution for normal incidence of tracks
for both directions of the pixel.
%
X corresponds to the direction of the long pixel side (400$\mu$m), Y
corresponds to the short direction (50$\mu$m). For the hit reconstruction first only the digital
information is used, i.e. the pixel with the largest signal above
threshold collected in a cone around the extrapolated track position
is taken as the hit pixel and its center is assumed to be the
reconstructed position. The digital
resolutions of pixel pitch divided by $\sqrt{12}$ in both directions,
i.e. 14.4\,$\mu$m and 114.5\,$\mu$m are smeared by a Gaussian
spread in the order of 5$\mu$m in y and 6$\mu$m in x, which is attributed to the
resolution of the track extrapolation, the detector noise ($\sim$380 e$^-$) and the
charge sharing behavior of the pixel cells. The structure visible
especially in fig.~\ref{resolution}(b), results from position
dependent response efficiencies which were subject to the
investigations above.
As will be shown below, charge sharing in 3D sensors is restricted to
a very narrow region of about $\sim$4$\mu$m (threshold dependent) at
the edge of a pixel. The improvement by using a charge-weighting algorithm (e.g. the so-called
$\eta$-algorithm~\cite{eta-algorithm}) compared to a purely digital readout is therefore limited for normally incident tracks.
In real vertex detectors charge sharing is usually purposely introduced by tilting the detectors
to improve spatial resolution through charge interpolation~\cite{ATLAS-pixel-paper}.
%
\paragraph*{Charge sharing}
Next we investigate the sharing of charge between pixel cells as a
function of the impact point of the track. This is studied by plotting
the mean value of the cluster size distribution, where a cluster is a
number of nearby hits above threshold. Also the fraction of charge
appearing in the pixel with the largest charge of the cluster (seed
pixel) is of interest.
These quantities are displayed as pixel maps in fig.~\ref{charge-sharging}.
The plots show most clearly the overall expected behavior.
On average the cluster size increases for inclined tracks.
Correspondingly the seed charge fraction decreases, modulated
by the position of the 3D electrodes and the resulting influence on the charge
collection efficiency.
\section{Conclusions}
3D active silicon sensors, assembled as pixel devices with the ATLAS
front end readout electronics have been tested and characterized for
the first time in a high energy (100 GeV) test beam using a 4-plane
microstrip beam telescope as reference. Devices with three 3D electrodes
under the area of a readout pixel
(50$\times$400 $\mu$m$^2$) were characterized. Hit detection efficiencies of
$95.9 \% \pm 0.1 \%$ ($99.9 \% \pm 0.1 \%$) for normal (15$^\circ$ inclined)
tracks have been obtained.
The inefficiencies for normally incident tracks have
been identified as being due the presence of the columnar 3D
electrodes. For inclined tracks the hit efficiency
approaches 100$\%$. The spatial distribution of efficiencies and the
mean cluster size as well as the fraction of charge seen by the seed pixel,
which characterize the charge sharing, have been measured.
\section*{Acknowledgments}
The authors would like to thank the CERN SPS staff
for their help during data taking.
%
\bibliographystyle{unsrt}
%

\end{document}